\documentclass[%
 reprint,
 amsmath,amssymb,
 aps,prd
floatfix,
showkeys
]{revtex4-2}
\usepackage{lmodern}
\usepackage{xcolor}
\usepackage[normalem]{ulem}
\usepackage{graphicx}
\usepackage{dcolumn}
\usepackage{bm}
\usepackage{hyperref}

\usepackage{float}
\usepackage[italicdiff]{physics}
\usepackage{tikz}
\usepackage{tikz-feynman}
\usepackage[caption = false]{subfig}
\usepackage{enumerate}
\usepackage{enumitem}
\usepackage{siunitx}
\usepackage{lipsum}
\usepackage{amsmath}
\usepackage{amsfonts}
\usepackage{amssymb}
\usepackage{amsthm}
\usepackage{mathtools}

\renewcommand{\a}{\alpha}								
\renewcommand{\b}{\beta}								
									
\renewcommand{\d}{\delta}								
\newcommand{\e}{\varepsilon}							
									
\newcommand{\D}{\Delta}									
\renewcommand{\epsilon}{\varepsilon}

\newcommand{\nn}{\nonumber}

\graphicspath{{plots/}}

\begin{document}
\title{Thermal corrections to Regge trajectories}

\author{R. Cádiz}
\affiliation{Department of Physics and Astronomy, Stony Brook University, Stony Brook, New York 11794-3800}

\author{M. Loewe}%
\affiliation{Facultad de Ingeniería, Universidad San Sebastián, Bellavista 7, Recoleta, Santiago, Chile.}
\affiliation{Centre for Theoretical and Mathematical Physics, and Department of Physics, University of Cape Town, Rondebosch 7700, South Africa}
\author{R. Zamora}
\affiliation{Instituto de Ciencias Básicas, Universidad Diego Portales, Casilla 298-V, Santiago, Chile.\\
Facultad de Ingeniería, Universidad San Sebastián, Bellavista 7, Recoleta, Santiago, Chile.}


\begin{abstract}
In this work, we investigate the behavior of Regge trajectories in the context of quantum field theory at finite temperature. For this purpose, we employ the $\lambda \phi^3$ model. We first compute the Regge trajectories at zero temperature, establishing a baseline for comparison. As a key novelty, we extended our analysis to finite temperature and derive an analytical expression for the Regge trajectories in this regime. We explore how temperature modifies the structure of the Regge trajectories and analyze the physical implications of these modifications, which arise from the resummation of thermal ladder diagrams. As an interesting consequence, we find the thermal evolution of masses belonging to different Regge trajectories.
\end{abstract}

\maketitle


\section{\label{sec:level1}Introduction}

The study of Regge trajectories, initially developed in the context of hadron scattering in the 1960s \cite{Regge:1959mz}, continues to provide profound insight into the analytic structure of scattering amplitudes and the behavior of quantum field theories (QFTs) at high energies. These trajectories link the spin of particles to their mass squared in a nearly linear relationship, encapsulating fundamental dynamics of hadronic resonances and offering a geometrical interpretation of scattering processes in strong interactions. Traditionally, Regge theory has been applied successfully in the context of zero temperature QFT \cite{PhysRevLett.7.394,PhysRevLett.8.41,Collins:1971ff,PhysRevD.63.054023,Irving:1977ea,PhysRevD.97.114001,Regge2,Regge3,Regge4,Regge5,Regge6,Regge7,Regge8,Regge9,Regge10,Ebert:2009ub}. However, finite temperature systems are of great interest in the context of QFT.  For example, different aspects concerning the behavior of hadronic parameters and the properties of the QCD phase diagram in the presence of thermal effects have been considered during the last years \cite{thermal1,paper1,thermal2,thermal9,paper2,thermal3,thermal4,paper3,thermal5,revisar2,thermal6,thermal7,thermal8,revisar1,thermal10,paper4}.

This paper explores the influence of thermal effects on Regge trajectories within the framework of a particular model in QFT.

This article is organized as follows. In Section \ref{sec2}, we will present the topic of Regge trajectories. Then, in Section \ref{sec3}, we will explore the Regge trajectories in the presence of temperature, and finally, in Section \ref{sec4}, we will present our conclusions.

\section{\label{sec2}Regge Trajectories: General Considerations}
\begin{figure}
\centering
\includegraphics[scale=0.4]{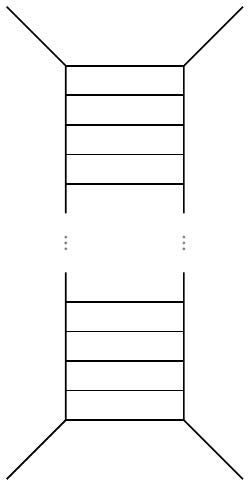}
\caption{Scattering process where “ladder" contributions are considered.}
\label{fig:LadderDiag}
\end{figure}

In this section, we aim to get the power-law behavior of scattering amplitudes, characteristic of Regge theory, by considering a high energy scattering process in the $s$ channel, when the $t$ channel has a fixed momentum. In particular, we are interested in the resummation of ladder diagrams as in Fig. \ref{fig:LadderDiag}.

\begin{figure}[hbtp]
\centering
\includegraphics[scale=0.4]{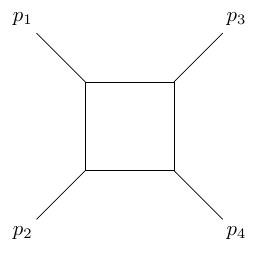}
\caption{Box diagram with circulating momentum $k$.}
\label{fig:BoxDiagram}
\end{figure}

Since we are interested in obtaining a thermal correction to the Regge trajectories, we begin our analysis by studying the zero temperature case, in the context of usual scalar field theory. In this sense, it is necessary to analyze the box diagram shown in Fig. \ref{fig:BoxDiagram}, which has the following expression 
\begin{equation}
i\mathcal{M}=\lambda^4\int\dfrac{\dd[4]{k}}{\qty(2\pi)^4}D(k)D(k+p_1)D(k+p_1+p_3)D(k-p_2),
\label{IntegralFull}
\end{equation}
with $D(k)$ the usual scalar propagator in Minkowski space, given by
\begin{equation}
D(k)=\dfrac{i}{k^2-m^2 +i\epsilon},
\end{equation}
and where $p_1, p_2$ are the incoming momenta and $p_3, p_4$ are the outgoing momenta. In order to compute Eq. (\ref{IntegralFull}), we introduce four Feynman parameters, which allows us to write,
\begin{widetext}
\begin{align}
i\mathcal{M}&= \lambda^4\int\dfrac{\dd[4]{k}}{(2\pi)^4}\qty(\dfrac{i}{k^2-m^2})\qty(\dfrac{i}{\qty(k+p_1)^2-m^2})\qty(\dfrac{i}{\qty(k+p_1-p_3)^2-m^2})\qty(\dfrac{i}{\qty(k-p_2)^2-m^2}) \nn\\[7pt]
&=\lambda^4 \int\dfrac{\dd[4]{k}}{(2\pi)^4}\int\dfrac{3!\d\qty(x+y+z+w-1)\dd{x}\dd{y}\dd{x}\dd{w}}{\qty[x\qty(k^2-m^2)+y\qty(\qty(k+p_1)^2-m^2)+z\qty(\qty(k+p_1-p_3)^2-m^2)+w\qty(\qty(k-p_2)^2-m^2)]^4}.
\end{align}
\end{widetext}

To continue the computation of this integral, we can manipulate the denominator to isolate the dependence $k$ by the usual completion of squares. After the corresponding algebra, we can perform the change of variables $k\to q-yp_1-z(p_1-p_3)+wp_2$, to get
\begin{align}
i\mathcal{M}&=3!\lambda^4\int_0^1 \dd{x}\dd{y}\dd{z}\dd{w} \delta(x+y+z+w-1) \nonumber \\
&\hspace*{1cm}\cdot\int \dfrac{d^4q}{(2\pi)^4} \frac{1}{(q^2-\Delta)^4},\label{Integral3}
\end{align}
where we have defined
\begin{align}
\D&=yws-zt\qty(1-z)+2z\qty(p_1-p_3)\qty(yp_1-wp_2)\nn\\
&\hspace{1cm}-m^2\qty[y+w-1-\qty(y-w)^2],
\label{DeltaZeroT}
\end{align}
with $s=\qty(p_1+p_2)^2$ and $t=\qty(p_1-p_3)^2$ the usual Mandelstam variables. The next step is to integrate over $q$, were we obtain,
\begin{equation}
\mathcal{M}=\dfrac{\lambda^4}{96\pi^2}\int\dfrac{\d\qty(x+y+z+w-1)\dd{x}\dd{y}\dd{z}\dd{w}}{\Delta^2}.
\end{equation}

In order to compute the remaining integral, we use Collins' technique \cite{Collins}, which exploits the fact that we are working in the regime when $s\to\infty$, which allows us to identify that the most relevant contribution in the denominator is given by the first term, while we can safely neglect the other terms containing $y$ and $w$. Therefore, and focusing only on the integrals over $y,w$,
\begin{align}
&\int_{0}^{\e}\dfrac{1}{\qty[yws-zt\qty(1-z)+m^2]^2}\dd{y}\dd{w}\nn\\[7pt]
&\hspace*{1cm}= \int_{0}^{\e}\dfrac{\e}{\qty[m^2-zt\qty(1-z)]\qty[sy\e-zt\qty(1-z)+m^2]}\dd{y} \nn\\[7pt]
&\hspace*{1cm}= \dfrac{1}{s\qty[m^2-zt\qty(1-z)]}\ln(\dfrac{s\e^2-zt\qty(1-z)+m^2}{m^2-zt\qty(1-z)}) \nn\\[7pt]
&\hspace*{1cm}\approx \dfrac{1}{m^2-zt\qty(1-z)}\cdot\dfrac{\ln s}{s}
\end{align}
\label{IntegralFeynman}
and therefore, the remaining integral looks like
\begin{equation}
    \mathcal{M}=\dfrac{\lambda^4}{96\pi^2}\qty(\dfrac{\ln s}{s})\int_{0}^{1}\int_{0}^{1}\dfrac{\d\qty(x+z-1)}{m^2-zt\qty(1-z)}\dd{x}\dd{z}.\label{8}
\end{equation}
The logarithm term clearly emerges form the asymptotic region associated to the leading term when $s \rightarrow \infty$.  Finally, by computing the integrals over $x$ and $y$, we arrive at
\begin{equation}
\mathcal{M}=\lambda^2K(t)\qty(\dfrac{\ln s}{s}),
\label{IntegralFinal}
\end{equation}
where we have defined
\begin{equation}
K(t)=\dfrac{\lambda^2}{24\pi^2\sqrt{t\qty(4m^2-t)}}\arctan(\sqrt{\dfrac{t}{4m^2-t}}).
\end{equation}

Having determined the analytical behavior of the box diagram, we now proceed to compute the Regge trajectory. To obtain the trajectory, we must proceed with the resummation of diagrams by stacking box diagrams to form a “staircase diagram'' as shown in Fig. \ref{fig:LadderDiag}. Taking into account each loop, this ladder diagram with n rungs in the $s \rightarrow \infty$ with $t$ fixed, can be expressed as
\begin{equation}
\mathcal{M}_n= \dfrac{\lambda^2}{s}\dfrac{\qty[K(t)\ln s]^{n-1}}{\qty(n-1)!}.
\label{TotalIntegral}
\end{equation}
Thus, summing an infinite series of ladder diagrams yields the asymptotic behavior of the resulting amplitude as
\begin{eqnarray}
A(s,t)&=&\sum_{n=1}^{\infty} \mathcal{M}_n \nonumber \\
&=& \frac{\lambda^2}{s} \sum_{n=1}^{\infty}\dfrac{\qty[K(t)\ln s]^{n-1}}{\qty(n-1)!} \nonumber \\
&=&\dfrac{\lambda^2}{s}\cdot e^{K(t)\ln s} \nonumber \\
&=&\lambda^2s^{\a(t)},
\end{eqnarray}
in which we can identify the Regge trajectory as
\begin{equation}
\a(t)=K(t)-1.
\end{equation}

\noindent
Notice that we need $t < 4m^2$, i.e. $t$ less than the s threshold for pair production, to have a real value for $K(t)$.


\section{\label{sec3}Finite Temperature Regge Trajectories}
In order to calculate the finite temperature case, we start from Eq. (\ref{Integral3}), and incorporate temperature effects using the imaginary time formalism in the standard way \cite{Bellac:2011kqa}, i.e.,
\begin{eqnarray}
\int \frac{d^4q}{(2\pi)^4}f(q) \rightarrow \frac{i}{\beta}\sum_{n} \int \frac{d^3q}{(2\pi)^3} f(\omega_n,\vb{q}),
\end{eqnarray}
where we have introduced the bosonic Matsubara frequencies $\omega_{n} = 2\pi n/\beta$ with $\beta=1/T$. Therefore, the expression for the box diagram at finite temperature is
\begin{eqnarray}
\mathcal{M}_{0,T}&=&3!\lambda^4\int_0^1 \dd{x}\dd{y}\dd{z}\dd{w} \delta(x+y+z+w-1) \nonumber \\
&\times&\frac{1}{\beta}\sum_{n} \int \frac{d^3q}{(2\pi)^3} \frac{1}{(\omega_n+\vb{q}^2+\Delta)^4},
\end{eqnarray}
where the subscripts $0$ and $T$ correspond to vacuum and finite temperature, respectively. Using the following identity
\begin{equation}
\frac{1}{(\omega_n+\vb{q}^2+\Delta)^4}=-\frac{1}{3!}\left(\frac{\partial}{\partial \Delta}\right)^3\frac{1}{(\omega_n+\vb{q}^2+\Delta)},
\end{equation}
we obtain
\begin{align}
\mathcal{M}_{0,T}&=-\lambda^4\int_0^1 \dd{x}\dd{y}\dd{z}\dd{w} \delta(x+y+z+w-1) \nonumber \\
&\hspace*{0.7cm}\times\left(\frac{\partial}{\partial \Delta}\right)^3\frac{1}{\beta}\sum_{n} \int \frac{d^3q}{(2\pi)^3} \frac{1}{(\omega_n+\vb{q}^2+\Delta)}
\end{align}
and performing the summation over Matsubara frequencies, we get

\begin{eqnarray}
\mathcal{M}_{0,T}&=&-\lambda^4\int_0^1 \dd{x}\dd{y}\dd{z}\dd{w} \delta(x+y+z+w-1) \nonumber \\
&\times&\left(\frac{\partial}{\partial \Delta}\right)^3 \int \frac{d^3q}{(2\pi)^3} \Biggl( \frac{1}{2(\vb{q}^2+\Delta)^{1/2}} \nonumber \\
&+& \frac{1}{(\vb{q}^2+\Delta)^{1/2}}e^{-\b\sqrt{\vb{q}^2+\D}}\Biggr) \nonumber \\
&\equiv& \mathcal{M}_{0}+\mathcal{M}_{T}, \label{20}
\end{eqnarray}
where 
\begin{align}
\mathcal{M}_{0}&=-\lambda^4\int_0^1 \dd{x}\dd{y}\dd{z}\dd{w} \delta(x+y+z+w-1) \nonumber  \\
&\hspace*{0.7cm}\times\left(\dfrac{\partial}{\partial \Delta}\right)^3 \int \frac{d^3q}{(2\pi)^3}  \dfrac{1}{2(\vb{q}^2+\Delta)^{1/2}}, 
\label{19}
\end{align}
and
\begin{align}
\mathcal{M}_{T}&=-\lambda^4\int_0^1 \dd{x}\dd{y}\dd{z}\dd{w} \delta(x+y+z+w-1) \nonumber \\
&\hspace*{0.7cm}\times\left(\dfrac{\partial}{\partial \Delta}\right)^3 \int \dfrac{d^3q}{(2\pi)^3}\dfrac{1}{(\vb{q}^2+\Delta)^{1/2}}e^{-\b\sqrt{\vb{q}^2+\D}}.\label{termica}
\end{align}
We note that $\mathcal{M}_{0}$ is temperature independent and corresponds to the term calculated in Eq. (\ref{IntegralFinal}). 
The temperature dependence is encoded in the term $\mathcal{M}_{T}$ being its calculation not trivial. In  order to handle this expression correctly, recalling that we are in the context of Regge trajectories, we take the limit $s\to\infty$ and, therefore, $\Delta\to\infty$. Using this limit and differentiating Eq. (\ref{termica}) with respect to $\Delta$, we obtain
\begin{widetext}
\begin{align}
\mathcal{M}_{T}&=-\lambda^4\int_0^1 \dd{x}\dd{y}\dd{z}\dd{w} \delta(x+y+z+w-1) \nonumber \\
            &\hspace*{0.7cm}\times\Biggl(-\int\dfrac{\dd[3]{q}}{(2\pi)^3}\qty[\frac{\beta ^3}{8 \left(\vb{q}^2+\Delta\right)^2}+\frac{3 \beta ^2}{4 \left(\vb{q}^2+\Delta\right)^{5/2}}+\frac{15 \beta }{8 \left(\vb{q}^2+\Delta\right)^3}+\frac{15}{8 \left(\vb{q}^2+\Delta\right)^{7/2}}]e^{-\b\sqrt{\vb{q}^2+\D}} \Biggr) \label{23}.
\end{align}
\end{widetext}
Since we are interested in the low temperature regime, we will focus on the first term of Eq. (\ref{23}). Therefore, integrating over the variable $q$, we get
 \begin{equation}
\mathcal{M}_{T}=\lambda^4\int_0^1 \dd{x}\dd{y}\dd{z}\dd{w} \delta(x+y+z+w-1)\dfrac{\b^3}{64 \pi \sqrt{\D}}e^{-\b \pi \sqrt{\D}}.
\end{equation}
To perform the calculation in Feynman parameters, we will use the Collins method employed in the previous section. That is, we need to integrate the following expression
\begin{eqnarray}
\mathcal{M}_{T}&=&\lambda^4\int \dd{x}\dd{y}\dd{z}\dd{w} \delta(x+y+z+w-1)\nonumber \\
&\times&\dfrac{\b^3}{64 \pi \sqrt{y s w+t z (1-z)+m^2)}}e^{-\b\sqrt{y s w+t z (1-z)+m^2}}.\nonumber \\
\end{eqnarray}
Integrating over the variables $y$,$w$ and $x$, we obtain
\begin{equation}
\mathcal{M}_{T}=-\lambda^4 \left(\frac{\ln s}{s} \right)\int_0^1 \dd{z} \frac{\beta^2e^{-\beta \sqrt{t z (1-z)+m}}}{32 \pi},
\end{equation}
let us note that this last integral over $dz$  cannot be performed analytically. 
Observe that the preceding equation has a form analogous to that found in Eq. (\ref{8}). Therefore, we can express it as follows
\begin{equation}
\mathcal{M}_{T}=\lambda^2B(t,\beta) \left(\frac{\ln s}{s}\right),
\end{equation}
where 
\begin{equation}
B(t,\beta)=-\lambda^2\int_0^1 \dd{z} \frac{\beta^2e^{-\beta \sqrt{t z (1-z)+m}}}{32 \pi}.\label{btb}
\end{equation}
Finally, adding the zero temperature result, we obtain
\begin{equation}
\mathcal{M}_{0,T}= \lambda^2(K(t)+B(t,\beta))\left(\frac{\ln s}{s}\right).
\end{equation}
Performing the analogous calculation to Section (\ref{sec2}), taking into account each loop, we obtain in this case
\begin{equation}
\mathcal{M}_{0,T,n}= \dfrac{\lambda^2}{s}\dfrac{\qty[(K(t)+B(t,\beta))\ln s]^{n-1}}{\qty(n-1)!}.
\end{equation}
Therefore, summing an infinite series of ladder diagrams, we get
\begin{eqnarray}
A(s,t,\beta)&=&\sum_{n=1}^{\infty} \mathcal{M}_{0,T,n} \nonumber \\
&=& \frac{\lambda^2}{s} \sum_{n=1}^{\infty}\dfrac{\qty[(K(t)+B(t,\beta))\ln s]^{n-1}}{\qty(n-1)!} \nonumber \\
&=&\dfrac{\lambda^2}{s}\cdot e^{(K(t)+B(t,\beta))\ln s} \nonumber \\
&=&\lambda^2s^{\a(t,\beta)},
\end{eqnarray}
which, in this case, allows us to identify the Regge trajectory with a thermal contribution of the form
\begin{equation}    \alpha(t,\beta)=K(t)-1+B(t,\beta).
\end{equation}
Regge trajectories can be expressed as a relation between the spin of hadronic states and their squared mass, of the form
\begin{equation}
\alpha(t)= \alpha_0+\alpha' t, \label{alpha}
\end{equation}
where, $\alpha_0$ is the intercept with the vertical axis, $\alpha'$ is the Regge slope, and $t$  is the Mandelstam variable, which is associated with the squared mass of the excited states that appear as resonances, expressed in units of $\text{GeV}^2$. In order to obtain our analytic expressions in the form of Eq. (\ref{alpha}), we perform a small $t$ expansion, Let us note that with this expansion we will be able to obtain an analytical expression for Eq. (\ref{btb}) since we will be able to perform the integration over $dz$, yielding
\begin{eqnarray}
\alpha(t,\beta)&=&\frac{\lambda^2}{96 \pi^2m^2}-\lambda^2 \beta^2 \frac{e^{-m \beta}}{32 \pi}\nonumber \\
&+&\lambda^2 t\left(\frac{1}{576 \pi^2 m^4}- \beta^3\frac{e^{-m \beta}}{384 \pi m }\right).\label{alfat}
\end{eqnarray}
We note that we have obtained an analytical expression for the Regge trajectories at finite temperature. In the literature \cite{Collins}, several trajectories are known, but only at zero temperature. Some of them are the following
\begin{eqnarray}
\alpha_\rho(t) &\approx& 0.5 + 0.9t \nonumber \\
\alpha_{K^*}(t) &\approx& 0.3 + 0.9t\nonumber \\
\alpha_\phi(t) &\approx& 0.1 + 0.9t \nonumber \\
\alpha_\pi(t) &\approx& 0.0 + 0.8t \label{tablatra}
\end{eqnarray}
In order to analyze how the Regge trajectories change with temperature, we plot the trajectories given by Eq.~(\ref{tablatra}) at different temperatures. To do so, we fit the values of $\lambda$ and $m$ in such a way that, at $T=0$, the trajectory reproduces the correct one.
 \begin{figure}[hbtp]
\centering
\includegraphics[scale=0.25]{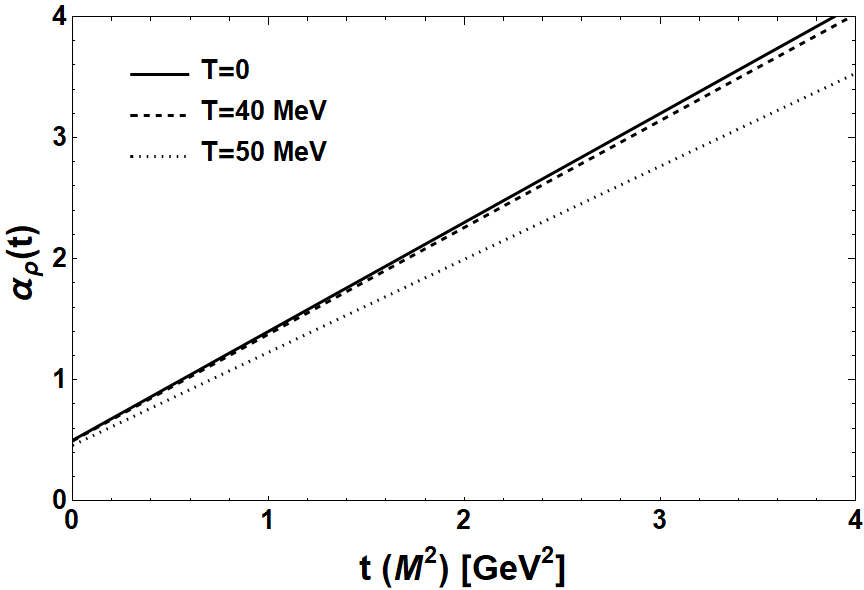}
\caption{$\alpha_\rho$(t) vs. t. The solid line corresponds to the Regge trajectory at $T = 0$, the dashed line to $T = 40$ MeV, and the dotted line to $T = 50$ MeV.}
\label{trayrho}
\end{figure}

 \begin{figure}[hbtp]
\centering
\includegraphics[scale=0.25]{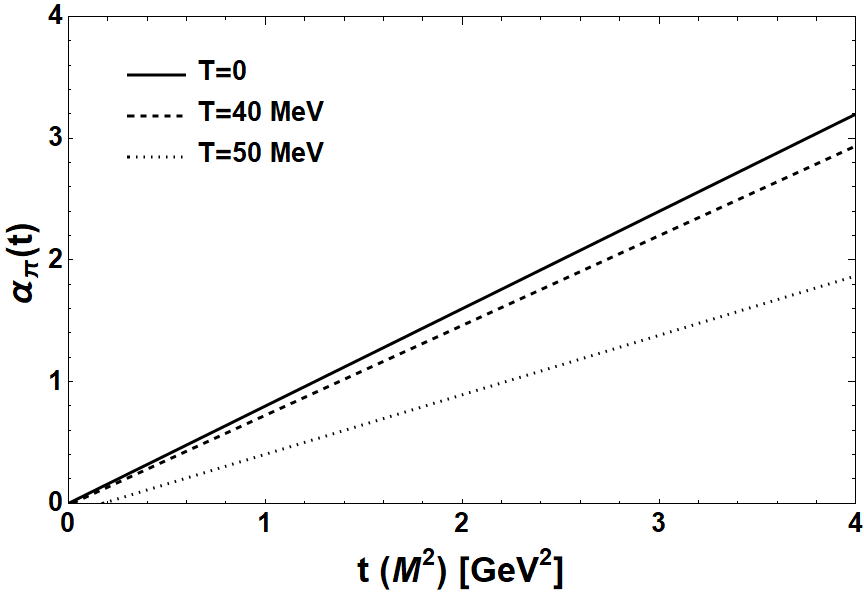}
\caption{$\alpha_\pi$(t) vs. t. The solid line corresponds to the Regge trajectory at $T = 0$, the dashed line to $T = 40$ MeV, and the dotted line to $T = 50$ MeV.}
\label{trayrho}
\end{figure}

 \begin{figure}[hbtp]
\centering
\includegraphics[scale=0.25]{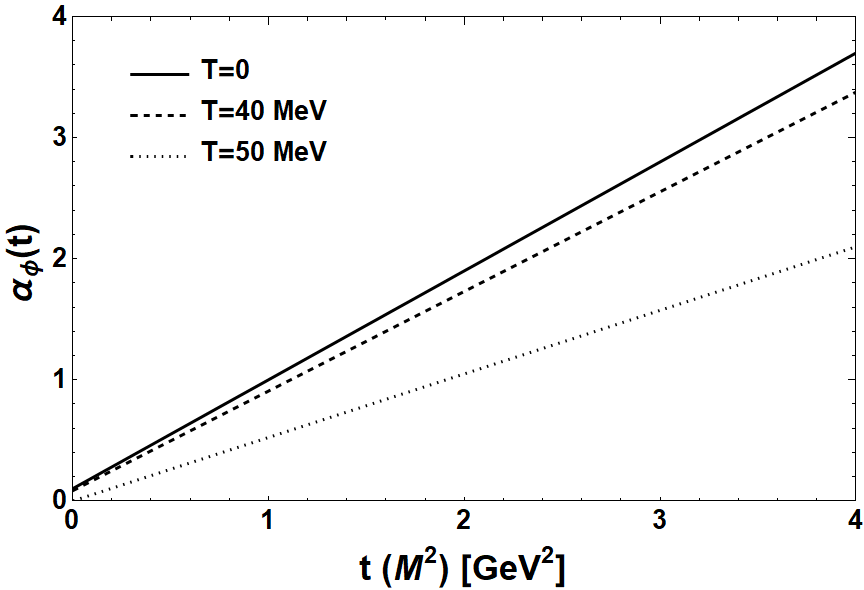}
\caption{$\alpha_\phi$(t) vs. t. The solid line corresponds to the Regge trajectory at $T = 0$, the dashed line to $T = 40$ MeV, and the dotted line to $T = 50$ MeV.}
\label{trayrho}
\end{figure}

 \begin{figure}[hbtp]
\centering
\includegraphics[scale=0.25]{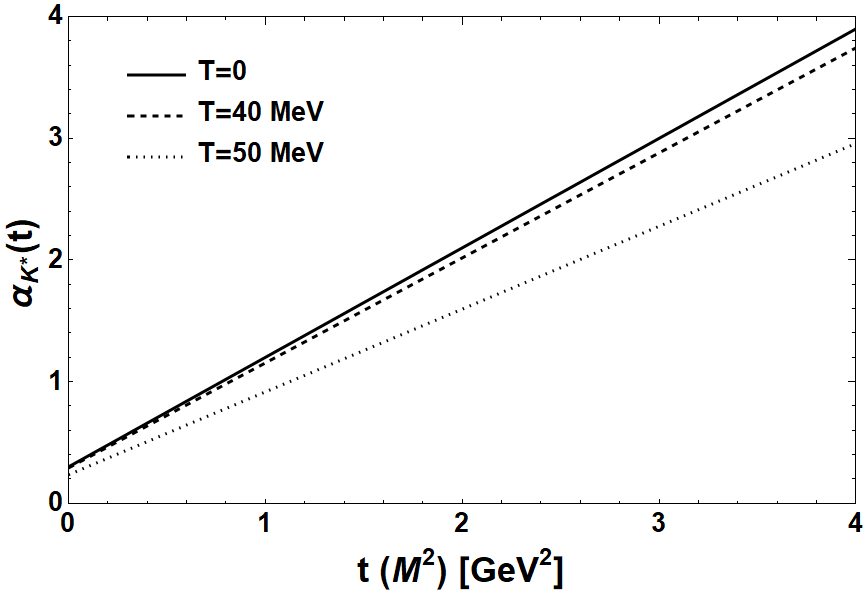}
\caption{$\alpha_{K^{*}}$(t) vs. t. The solid line corresponds to the Regge trajectory at $T = 0$, the dashed line to $T = 40$ MeV, and the dotted line to $T = 50$ MeV.}
\label{trayrho}
\end{figure}
\begin{figure}[hbtp]
\centering
\includegraphics[scale=0.37]{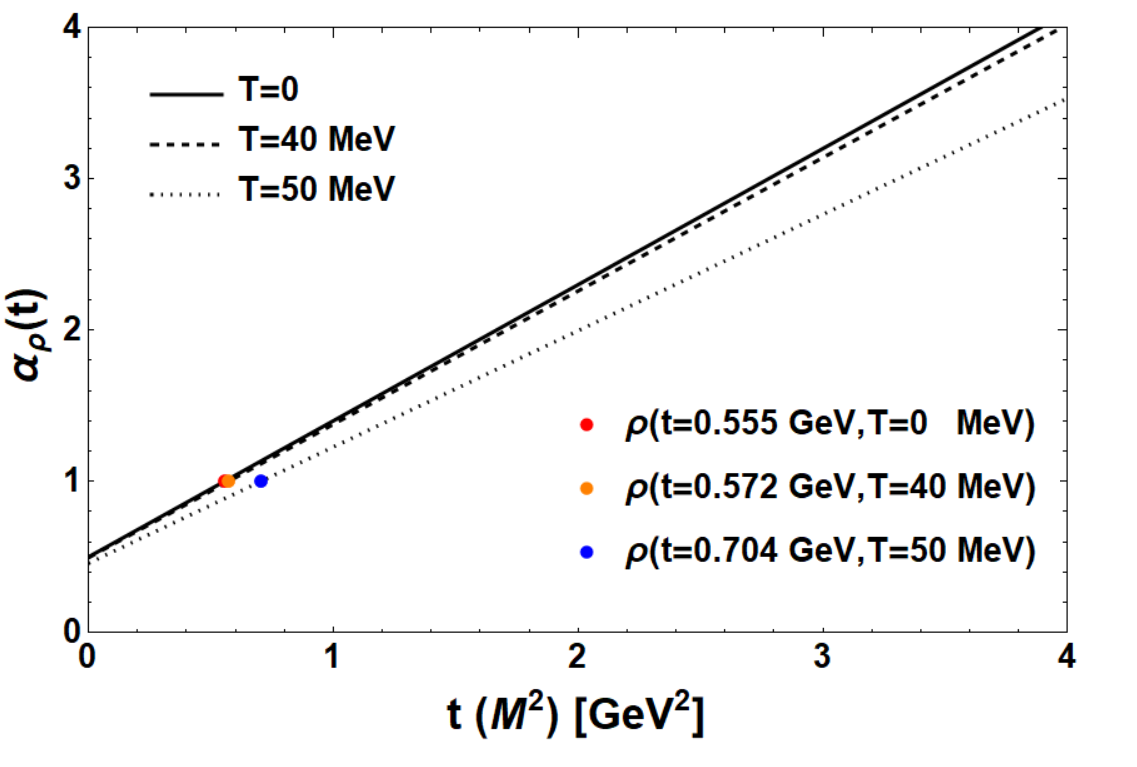}
\caption{$\alpha_{\rho}(t)$ vs. $t$. The solid line corresponds to the Regge trajectory at $T = 0$, the dashed line to $T = 40$ MeV, and the dotted line to $T = 50$ MeV. The red point indicates the squared mass of the $\rho$ meson in the vacuum case, the orange point corresponds to the $\rho$ squared mass at $T = 40$ MeV. and the blue point corresponds to the $\rho$ squared mass at $T = 50$ MeV.}
\label{grafico1}
\end{figure}
\begin{figure}[hbtp]
\centering
\includegraphics[scale=0.28]{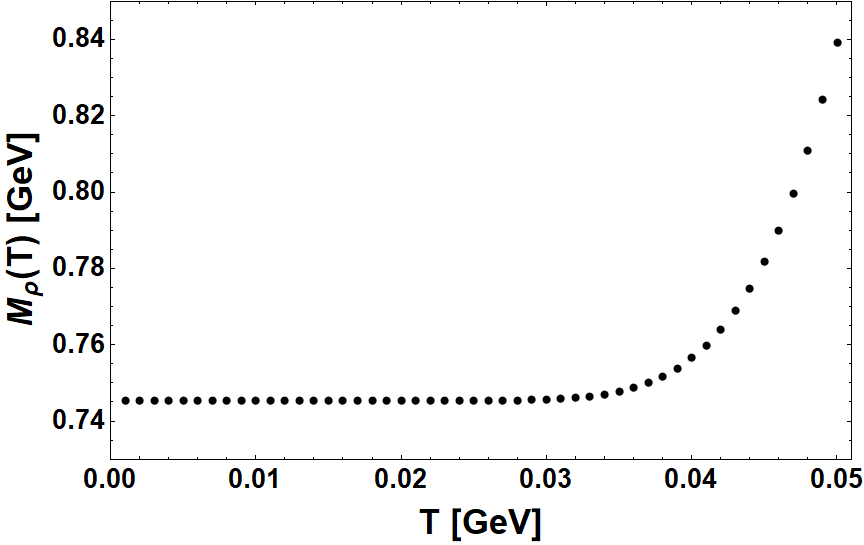}
\caption{$M_\rho$(T) vs. T}
\label{graficomasarho}
\end{figure}

The plots reveal that the Regge slope decreases with increasing temperature. Another interesting piece of information that can be extracted from the plots is the particle mass. For example, if we consider the trajectory of the $\rho$ meson, we can obtain the value of its mass at different temperatures, as shown in Fig. \ref{grafico1}. From Fig. \ref{grafico1},
we can extract the mass of the $\rho$ meson in the zero temperature case as approximately $0.744$~GeV, while at $T = 40$~MeV the $\rho$ meson mass is $0.756$~GeV and at $T = 50$~MeV the $\rho$ meson mass is $0.839$~GeV, we note that for small temperature values, there is no significant variation in the mass of the $\rho$ meson. However, as the temperature increases, the mass clearly begins to rise, in agreement with theoretical results Ref.~\cite{Dominguez:1992dw} where an analysis of the temperature evolution of the rho-meson mass was implemented. This was done in the context of unitarity restrictions, together with the virial expansion, applied to $\pi$-$\pi$ scattering lengths in the chiral expansion. It was shown in this work that the growing behavior of the width $\Gamma (T)$, which turns out to be the crucial order parameter for the deconfinement phase transition, occurs together with a growing behavior of the mass $m_{\rho}(T)$. Here we find a similar behavior for the rho mass evolution, in spite of the fact that our model is extremely simple. This is basically related to the first observation: the slope of the Regge trajectories decreases as function of temperature. Moreover, we can carry out the same procedure for the $\rho$ meson using several values of $T$, in order to obtain multiple Regge slopes and extract the mass of the $\rho$ as a function of temperature, resulting in Fig.~\ref{graficomasarho}.

\section{\label{sec4}Conclusions}

We have analytically derived the thermal corrections to Regge trajectories within a $\lambda \phi^3$ quantum field theory framework by resumming ladder diagrams using the imaginary-time formalism. Our results show that finite temperature induces a reduction in the Regge slope, which leads to observable effects such as the increase of meson masses with temperature. In particular, we analyze the behavior of the $\rho$ meson and find that its mass exhibits a mild increase at low temperatures, followed by a more rapid growth as the temperature increases.  This increase in mass aligns with theoretical predictions in thermal QFT~\cite{Dominguez:1992dw,mesonsigma}, and supports the idea that the excitation spectrum of hadrons is modified in the presence of a thermal bath.

\begin{acknowledgments}
M.L., and R.Z. acknowledge support from ANID/CONICYT FONDECYT Regular (Chile) under Grants No. 1220035 and 1241436.
\end{acknowledgments}

\appendix
\bibliography{bibliography}
\end{document}